\UseRawInputEncoding
\documentclass[twocolumn,superscriptaddress,nobibnotes,nofootinbib,floatfix,aps,prl,citeautoscript,reprint,longbibliography]{revtex4-1}

\usepackage{graphicx}
\usepackage{epsfig}
\usepackage{subfigure}
\usepackage{color}
\usepackage{latexsym}
\usepackage{amsmath,bm,multirow}
\usepackage{amsfonts}
\usepackage{dcolumn}           
\usepackage{natbib}
\usepackage{physics}

\usepackage[draft]{changes} 
\usepackage[normalem]{ulem}
\newcommand{\northwesternCHEM}{Department of Chemistry, Northwestern University, Evanston, IL 60208, USA}
\newcommand{\nw}{Department of Materials Science and Engineering, Northwestern University, Evanston, IL 60208, USA}

\usepackage[draft]{changes} 
\usepackage[normalem]{ulem}

\usepackage{soul}

\begin{document}

\title{Accelerated Discovery of a Large Family of Quaternary Chalcogenides with Very Low Lattice Thermal Conductivity}

\author{Koushik Pal}
\email{koushik.pal.physics@gmail.com}
\affiliation{\nw}

\author{Yi Xia}
\affiliation{\nw}

\author{Jiahong Shen}
\affiliation{\nw}

\author{Jiangang He}
\affiliation{\nw}

\author{Yubo Luo}
\affiliation{\northwesternCHEM}

\author{Mercouri G. Kanatzidis}
\affiliation{\northwesternCHEM}

\author{Chris Wolverton}
\email{c-wolverton@northwestern.edu}
\affiliation{\nw}



\begin{abstract}
The development of efficient thermal energy management devices such as thermoelectrics, barrier coatings, and thermal data-storage disks often relies on compounds that possess very low lattice thermal conductivity ($\kappa_l$). Here, we present the computational prediction of a large family of 628 thermodynamically stable quaternary chalcogenides, AMM'Q$_3$ (A = alkali/alkaline earth/post-transition metals; M/M' = transition metals, lanthanides; Q = chalcogens) using high-throughput density functional theory (DFT) calculations. We validate the presence of low-$\kappa_l$ in this family of materials by calculating $\kappa_l$ of several predicted stable compounds using the Peierls-Boltzmann transport equation within a first-principles DFT framework. Our analysis reveals that the low-$\kappa_l$ in the AMM'Q$_3$ compounds originates from the presence of either a strong lattice anharmonicity that enhances the phonon scatterings or rattlers cations that lead to multiple scattering channels in their crystal structures. Our predictions suggest new experimental research opportunities in the synthesis and characterization of these stable, low-$\kappa_l$ compounds.
\end{abstract}

\maketitle

\section{Introduction} 
 
An important focus in materials science research has been to discover novel materials with properties that might hold the keys to solving the most pressing problems in renewable energy, energy harvesting, or semiconductor power electronics. The augmentation of new materials discovery and the prediction of their properties have been accelerated by the advent of advanced computer algorithms coupled with high-throughput (HT) screening methods\cite{curtarolo2013high,mounet2018two,saal2013materials,kirklin2015open, jain2013commentary, jain2016computational, meng2009first, greeley2006computational, gautier2015prediction, zakutayev2013theoretical} using accurate quantum mechanical calculations based on density functional theory (DFT). In the recent past, several computational predictions have led to the successful synthesis of new solid-state compounds in a variety of chemistry and structure types  in the family of half-Heuslers\cite{gautier2015prediction, zakutayev2013theoretical}, double half-Heuslers\cite{anand2019double}, electrides\cite{wang2017exploration}, AB$_2$X$_4$ based chalcogendes\cite{xi2018discovery}, and rocksalt based compounds\cite{hao2014theoretical}. 

Crystalline solids with extreme thermal transport properties are technologically important for the efficient management of thermal energy\cite{samanta2020intrinsically}. While materials with high lattice thermal conductivity ($\kappa_l$) are used in microelectronic devices for heat dissipation, materials with low $\kappa_l$ are used in thermal barrier coatings\cite{darolia2013thermal}, thermal data-storage devices\cite{matsunaga2011phase}, and high-performance thermoelectrics (TEs)\cite{zhao2016ultrahigh,biswas2012high} which can convert heat into electrical energy.  The conversion efficiency of the TEs is defined by the figure-of-merit, ZT=$S^2\sigma/(\kappa_l + \kappa_e)$, where S, $\sigma$,  $\kappa_e$ and $\kappa_l$ are the Seebeck coefficient, electrical conductivity, electronic thermal conductivity, and lattice thermal conductivity, respectively. Engineering the electronic band structures of crystalline compounds that already possess low $\kappa_l$ has emerged to be a very popular strategy to increase the ZT. Therefore, crystalline semiconductors with intrinsically low-$\kappa_l$ are highly sought after in thermoelectrics and other thermal energy management devices. 

In pursuit of finding new low-$\kappa_l$ materials, different classes of crystalline compounds like the perovskites\cite{van2016high}, half-Heuslers\cite{carrete2014finding}, full-Heuslers\cite{he2016ultralow}, and double half-Heuslers\cite{anand2019double}, have been explored using HT computational methods, and subsequently, some of the predicted compounds were experimentally synthesized successfully\cite{gautier2015prediction, zakutayev2013theoretical, anand2019double}. Despite the existence of a large number of crystalline compounds in various materials databases e.g., the Inorganic Crystal Structure Database (ICSD)\cite{belsky2002new},  Open Quantum Materials Database (OQMD)\cite{saal2013materials,kirklin2015open}, Materials Project\cite{jain2013commentary}, and Aflowlib\cite{curtarolo2012aflowlib}, it is quite important to look for novel stable and metastable compounds which might exhibit exciting physical and chemical properties. In this work, we present the computational discovery of a large number of stable (628) and low-energy metastable  (852) quaternary chalcogenides AMM'Q$_3$ (A = alkali, alkaline earth, post-transition metals; M/M' = transition metals, lanthanides; Q = chalcogens) that span a huge chemical space across the periodic table.  Our results are based on reliable, accurate and robust HT-DFT calculations where (a) we generated initial AMM'Q$_3$ compositions following the experimentally known AMM'Q$_3$ compounds formation criteria, (b) calculated their energetics in all known seven crystallographic prototypes that are found in this materials family, and (c)  performed thermodynamic phase stability analysis of these compositions  against all possible competing phases that are present in the OQMD.

\begin{figure*}
\centering
\includegraphics[height=13cm, width=17cm, trim={0cm  0cm 0cm 0cm}, clip]{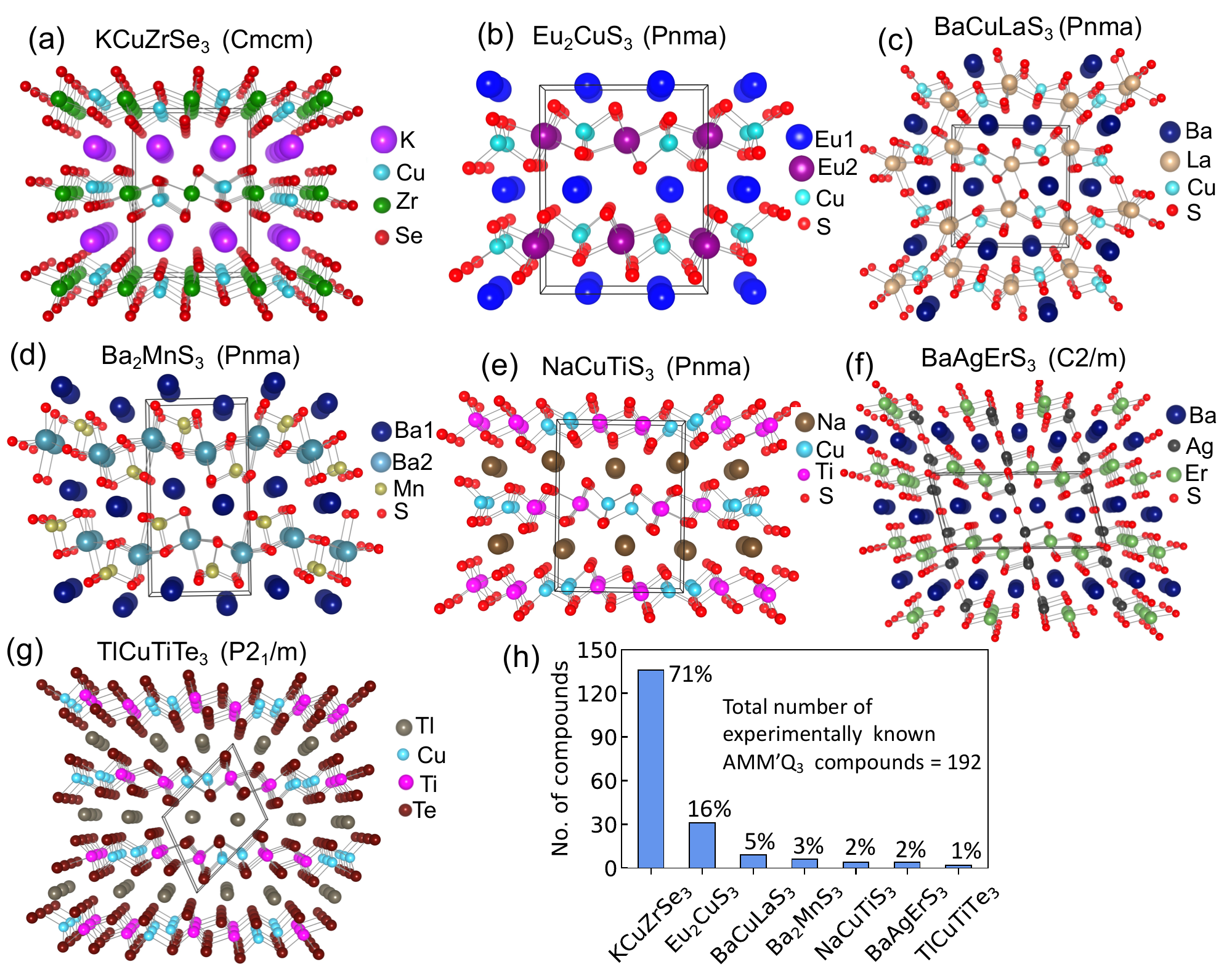}
\caption{\textbf{Crystallographic prototypes of the AMM'Q$_3$ compounds.} (a-g) Seven structural prototypes in the family of experimentally known AMM'Q$_3$ compounds. Most of them have layered crystal structures where the layers of A$^{m+}$ cations are sandwiched between the covalently bonded layers of [MM'Q$_3$]$^{m-}$ anion sublattice, and interact with the later through electrostatic interactions. The black solid lines in each figure indicate the conventional unit cell of each compound. (h) Distributions of the experimentally known AMM'Q$_3$ compounds in the seven structure types.}
\end{figure*}

About 192 quaternary chalcogenides (see Supplementary Information (SI) for a complete list) with the generic formula AMM'Q$_3$ have been synthesized experimentally\cite{koscielski2012structural,prakash2015syntheses,ruseikina2017crystal, ruseikina2019synthesis, ruseikina2019crystal, sikerina2007crystal, maier2016crystal, azarapin2020synthesis, strobel2006three} which reveal that these compounds possess rich chemistries and structure types like the perovskites and Heusler compounds. Koscielski et al.\cite{koscielski2012structural} noted that these known AMM'Q$_3$ compounds contain no Q-Q bonds and the elements (A, M, M', Q) balance  the  charge in their crystal structures  with their expected formal oxidation states, making them charge balanced. The AMM'Q$_3$ compounds are further classified into three categories depending on the nominal oxidation states of the three cations A, M, and M', namely:  
\begin{itemize}
    \item Type-I (A$^{1+}$M$^{1+}$M'$^{4+}$Q$_3$)
    \item Type-II (A$^{2+}$M$^{1+}$M'$^{3+}$Q$_3$)
    \item Type-III (A$^{1+}$M$^{2+}$M'$^{3+}$Q$_3$),
\end{itemize}

where the oxidation states of the cations are indicated with the superscripts. In all cases, we assume that the chalcogen atoms Q (S, Se, and Te) have a nominal 2- charge.  Examining the experimentally known AMM'Q$_3$ compounds (see SI Figure S1), we observe the following: 
(a) The A site in these compounds is always occupied by alkali, alkaline earth, or post-transition metals with the only exception of Eu (in a 2+ charge state) which occupies the A-site of some of the Type-II compounds.
(b) Whereas only transition metals occupy the M site, the M' site can be filled either by the transition metals, lanthanides, or actinides. 
(c) No observed AMM'Q$_3$ compound contains more than one alkali, alkaline earth, or post-transition metals.
As detailed later, we will use these criteria in designing our HT workflow for generating the initial crystal structures through prototype decoration. Although the crystal chemistries of these compounds have been characterized somewhat in detail\cite{koscielski2012structural,prakash2015syntheses,ruseikina2017crystal, ruseikina2019synthesis, ruseikina2019crystal, sikerina2007crystal, maier2016crystal, azarapin2020synthesis, strobel2006three}, their properties have remained largely unexplored. Recently, it was shown experimentally\cite{hao2019design} and theoretically\cite{hao2019design, pal2019intrinsically, pal2019unraveling, pal2019high} that many known semiconducting compounds in this crystal family exhibit ultralow $\kappa_l$. In addition, some compounds are shown to possess electronic bands favorable to support high thermoelectric performance\cite{hao2019design, pal2019unraveling, pal2019high}.

\begin{figure*}
\centering
\includegraphics[height=7cm, width=15cm, trim={0cm  0cm 0cm 0cm}, clip]{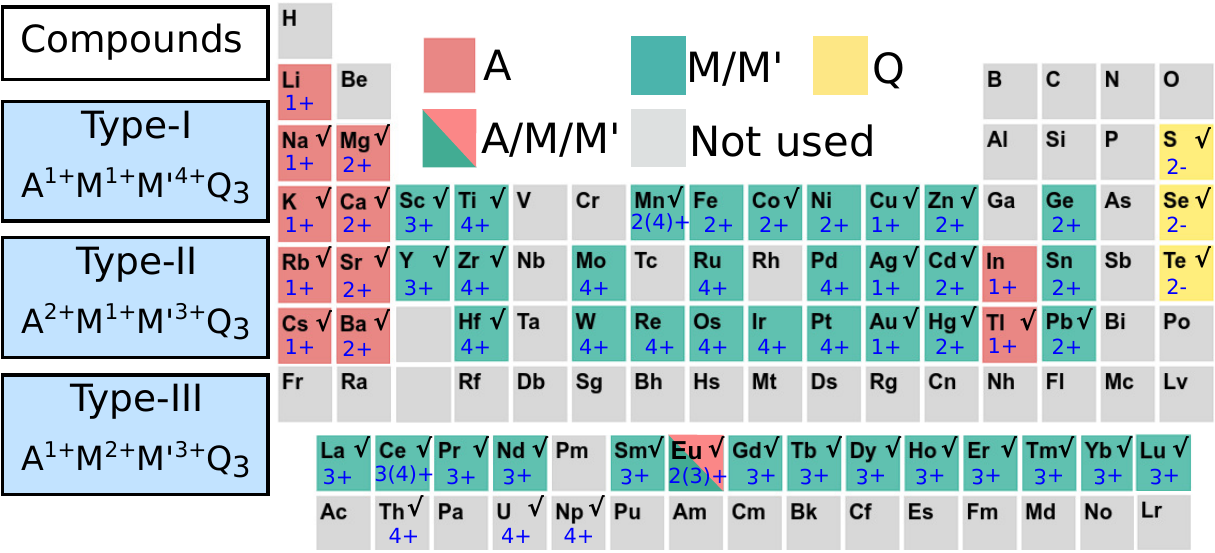}
\caption{\textbf{Elemental distributions of the experimentally known and theoretically decorated AMM'Q$_3$ compounds}. The AMM'Q$_3$ compounds are categorized into three types depending on the nominal oxidation states of the cations: Type-I (A$^{1+}$M$^{1+}$M'$^{4+}$Q$_3$), Type-II (A$^{2+}$M$^{1+}$M'$^{3+}$Q$_3$) and Type-III (A$^{1+}$M$^{2+}$M'$^{3+}$Q$_3$). The color-coded periodic table indicates the group of elements (whose oxidation states are written in blue color) that are used for prototype decorations in our HT-DFT calculations. Elements that are denoted with the check marks constitute the experimentally known AMM'Q$_3$ compounds. Salmon, cyan and yellow colors represent elements that occupy A, M/M' and Q sites, respectively. The grey color represents the elements that were not used for prototype decoration. Eu is the only element that occupies both A and M' site. All experimental and decorated compounds are charge-balanced.}
\end{figure*}

From the distribution of the elements forming the known AMM'Q$_3$ compounds, we see that all the cations (A, M, M') are in their common oxidation states of 1+, 2+, 3+ or 4+ coming from the chemical groups (i.e., alkali, alkaline earth, transition, post-transition metals, lanthanides, actinides) that span a large part of the periodic table. Yet, although a large number (192) of these compounds have been reported experimentally, this number is small compared to the vast number of possible compounds that can be obtained based on charge-balanced combinatorial substitutions of the elements in a prototype crystal structure of AMM'Q$_3$.  Performing this combinatorial exercise experimentally would require a massive amount of resources and time to discover new AMM'Q$_3$ compounds. However, computational screening can be very helpful in narrowing down the search space of the target compounds that would have higher chances of synthesizability in the laboratory\cite{curtarolo2013high, greeley2006computational, gautier2015prediction, zakutayev2013theoretical, anand2019double, hautier2012computer, sun2016thermodynamic, aykol2018thermodynamic, jain2011high, ward2017atomistic, haastrup2018computational}. Here, we have performed HT-DFT calculations followed by accurate ground-state phase stability analysis, and suggest (T = 0 K) thermodynamically stable and metastable AMM'Q$_3$ compounds for experimental synthesis and exploration of their properties. Our calculations of the thermal transport properties of some of the predicted stable compounds using the Peierls-Boltzmann transport equation (PBTE) show that these compounds exhibit innate low-$\kappa_l$ due to the presence of strong lattice anharmonicity or rattler cations.

\section{Results}

\subsection{Structural prototypes}
The experimentally known AMM'Q$_3$ compounds crystallize in seven structure types\cite{koscielski2012structural,prakash2015syntheses,ruseikina2017crystal, ruseikina2019synthesis, ruseikina2019crystal, sikerina2007crystal, maier2016crystal, azarapin2020synthesis, strobel2006three}: KCuZrSe$_3$ (space group (SG): Cmcm, \#63), Eu$_2$CuS$_3$ (SG: Pnma, \#62), BaCuLaS$_3$ (SG: Pnma, \#62), Ba$_2$MnS$_3$ (SG: Pnma, \#62),  NaCuTiS$_3$ (SG: Pnma, \#62), BaAgErS$_3$ (SG: C2/m, \#12) and TlCuTiTe$_3$ (SG: P2$_1$/m, \#11). All these structure types are visualized in their extended unit cells in Figure 1, where the conventional unit cell is outlined with black lines. Among these,  five structure types (KCuZrSe$_3$, Eu$_2$CuS$_3$, Ba$_2$MnS$_3$, NaCuTiS$_3$, TlCuTiTe$_3$) are layered where the rows of A$^{m+}$ cations stack alternatively with the [MM'Q$_3$]$^{m-}$ layers and interact through electrostatic interactions\cite{pal2019high, pal2019intrinsically}.   The strength of the interactions vary with the charges on the cations (i.e., m+) and induce modifications in the structures as well as in their properties\cite{pal2019unraveling}. When the interactions between the layers increase significantly, the atoms from the neighboring [MM'Q$_3$]$^{m-}$ layers interact, giving rise to the three-dimensional channel structures (BaCuLaS$_3$ and BaAgErS$_3$). In BaCuLaS$_3$ (Figure 1c) and BaAgErS$_3$ (Figure 1f), two rows of the A-site cations occupy the empty spaces inside the channels formed by the M, M', and Q atoms. Figure 1h shows that 71\% of the known AMM'Q$_3$ compounds crystallize in the KCuZrSe$_3$ structure followed by 16\% of the compounds crystallizing in the Eu$_2$CuS$_3$ structure type. The rest of the known AMM'Q$_3$ compounds (13 \%) crystallize in the other five structure types.  It is worth noting that in Eu$_2$CuS$_3$, the Eu atoms have mixed oxidation states (Eu$^{2+}$CuEu$^{3+}$S$_3$) and sit in two different sites in the crystal structure. Similarly, the Ba atoms in Ba$_2$MnS$_3$ also occupy two different sites. We have used all these structure types in our HT-DFT design and discovery of new AMM'Q$_3$ compounds. We note that the KCuZrSe3 and TlCuTiTe$_3$ structure types have 12 atoms in their primitive unit cells and the rest of the structure types have 24 atoms.

\begin{figure*}
\centering
\includegraphics[scale=0.5, trim={0cm  0cm 0cm 0cm}, clip]{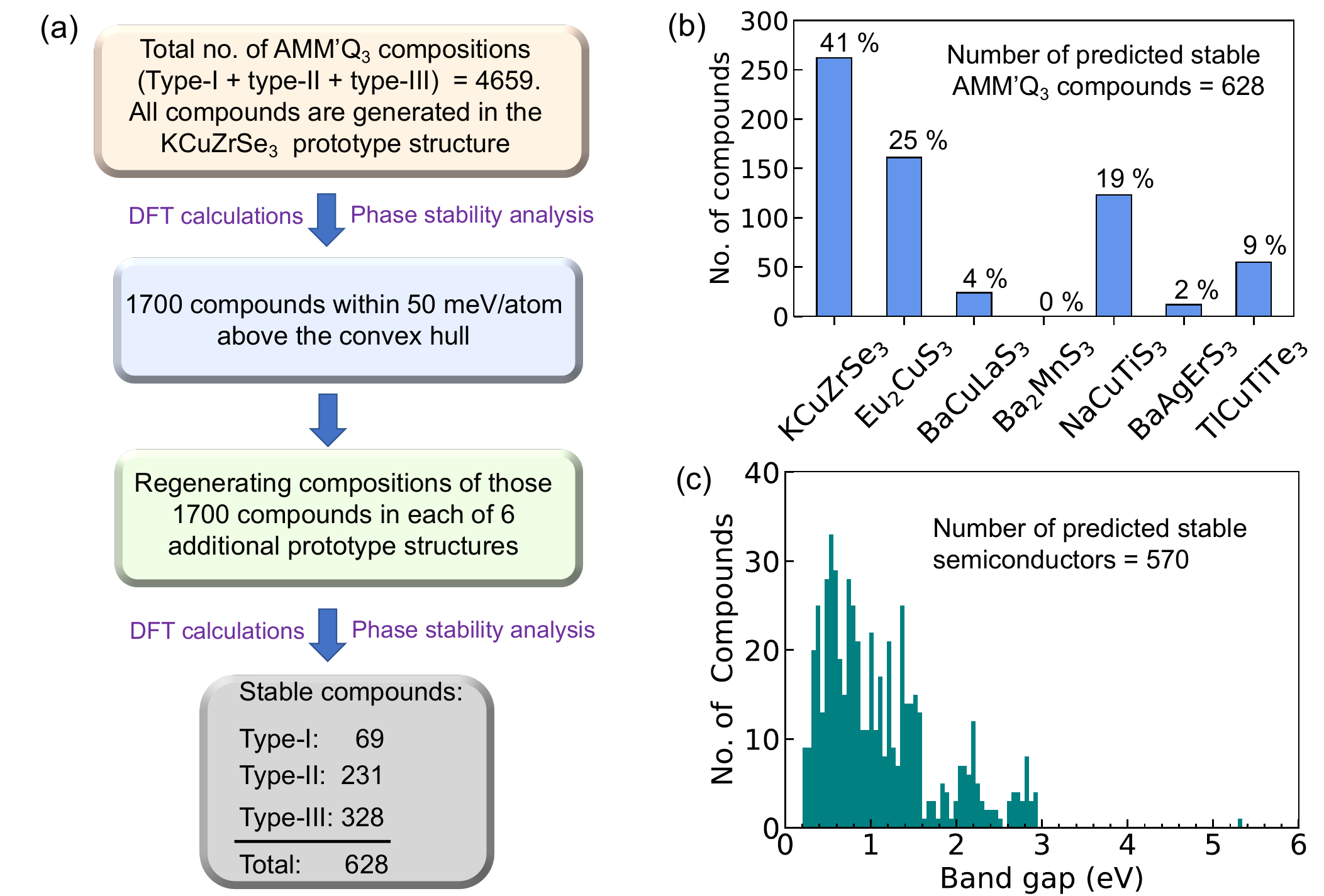}
\caption{\textbf{A brief workflow and summary of high-throughout DFT calculations}. (a) A schematic flow-chart of the HT-DFT calculations for the discovery of new  AMM'Q$_3$ compounds.  (b) The distribution of 628 predicted stable compounds in the 7 structure types. (c) 570 out of 628 compounds possess finite band gaps.  Histogram plot for the DFT-calculated band gaps of these 570 semiconductors.}
\end{figure*}

\subsection{Materials design strategy}
The discovery of new compounds through HT-DFT method often starts with the decoration of prototype crystal structures with chemically similar elements from the periodic table to generate their initial crystal structures. DFT calculations are then performed on these newly decorated compounds followed by rigorous thermodynamic phase stability analysis to screen for stable and metastable compounds. Rather than generating the input crystal structures in a brute-force manner by substituting every element of the periodic table at all atomic sites in the prototype structures, in this work we restricted ourselves by the following screening criteria which are derived by examining the experimentally known AMM'Q$_3$ compounds: 
(a) We substitute alkali, alkaline earth, or post-transition metal elements at the A site. The M and M' sites are populated with the transition metals and lanthanides. Three chalcogens i.e., S, Se, and Te are substituted at the Q site. 
(b) We choose A-site cations with nominal oxidation states of 1+ or 2+, and M/M'-site cations with nominal oxidation states of 1+, 2+, 3+, or 4+.  The elements that are chosen for substitutions at the A, M, M' and Q sites along with their oxidation states are shown in Figure 2. 
(c) We only consider compound compositions that are charge-balanced.
(d) We exclude any radioactive elements during the prototype decorations although some of the known AMM'Q$_3$ compounds contain them. 
 Adhering to these pre-conditions helps us narrow down our search space of compound exploration, reduces the computational cost, and most importantly it increases the success rates of stable compound prediction through HT-DFT calculations which will be evident later.

First, we generate the crystal structures of 4659 AMM'Q$_3$ compositions (see Table I) using the KCuZrSe$_3$ structure as (i) it is the most prevalent structure type in this family and (ii) all experimentally known AMM'Q$_3$ compounds have low energies (within 50 meV/atom above the convex hull) in this structure type. After performing DFT calculations for all these compositions followed by T=0 K phase stability analysis, we kept only those compounds ( $\sim$ 1700) that have an  energy within 50 meV/atom of the ground state convex hull and discarded the rest from our search space. The DFT relaxed structures of all these 1700 compounds retain the KCuZrSe$_3$ structure type. In the next step, we take these 1700 compositions and regenerate their crystal structures in each of six additional structure types to perform DFT calculations of 6 $\times$ 1,700 = 10,200 AMM'Q$_3$ compounds.  Next, we perform T = 0K thermodynamic phase stability analysis of those 1700 compositions considering all seven structure types and their competing phases that are available in the OQMD. In the final step, we obtain 628 thermodynamically stable and 852 metastable hitherto unknown AMM'Q$_3$ compounds after performing a total number of 4,659 + 10,200 = 14,859 DFT calculations. The stable 628 compounds include 69 Type-I, 231 Type-II, and 328 Type-III compounds, and among them, a total number of 570 compounds possess finite band gaps.  A schematic of the HT-DFT flowchart is shown in Figure 3a, and a summary is given in Table I.

\begin{figure*}
\centering
\includegraphics[scale=0.5, trim={0cm  1cm 0cm 0cm}, clip]{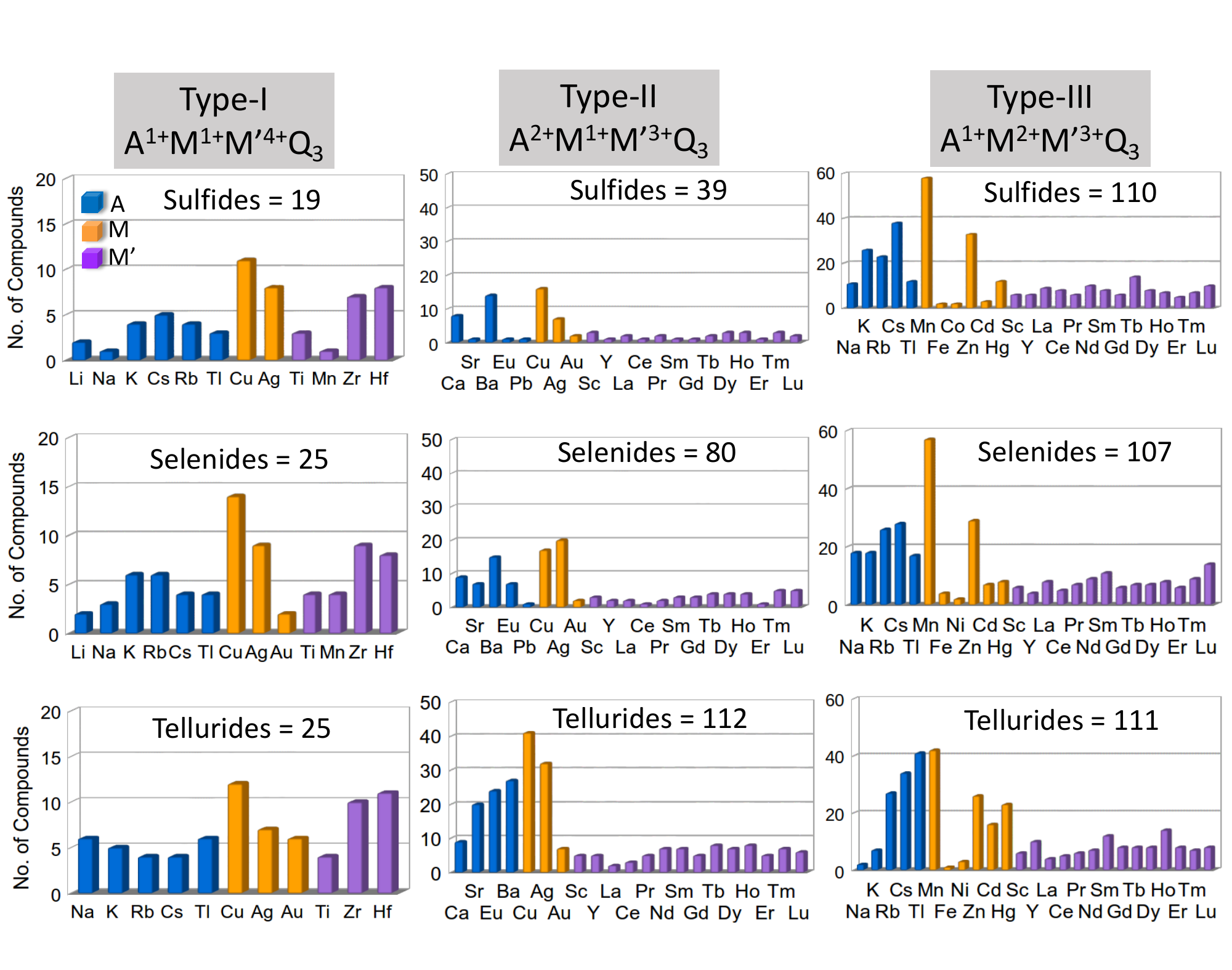}
\caption{\textbf{Elemental distributions of the predicted stable AMM'Q$_3$ compounds}. Distributions of the cations (A, M, M') forming the 628  predicted stable AMM'Q$_3$ compounds in the three categories: Type-I, Type-II, and Type-III. The bar corresponding to an element represents the number of stable compounds containing it. Total number of  sulfides, selenides and tellurides are mentioned in each panel. These compounds exclude the experimentally known 192 AMM'Q$_3$ compounds.}
\end{figure*}

\subsection{Phase stability analysis}
We now present a detailed analysis of the T= 0 K ground state phase stability of all (known and predicted) AMM'Q$_3$ compounds. We begin our assessment with the phase stability analysis of the experimentally known AMM'Q$_3$.  Out of 192 known compounds, we find only 119 compounds (Type-I: 39, Type-II: 30, and Type-III: 50) in OQMD before we performed any new calculations from this work. We designate these 119 compounds as Set-I.  As the initial crystal structures of the experimentally known compounds in OQMD mostly come from the ICSD, the DFT calculations of Set-I compounds in OQMD were performed based on their experimental crystal structures taken from the ICSD. We  designate the rest of the known 192-119 = 73 compounds (Type-I: 1, Type-II: 60, and Type-III: 12) as Set-II, which were not present in OQMD due to (a) the absence of their structures in the ICSD and (b) no previous HT-DFT calculations were performed based on the prototype decorations in this AMM'Q$_3$ family. However, our HT-DFT calculations of all the decorated AMM'Q$_3$ compositions include the experimentally known AMM'Q$_3$ compounds in Set-II.  Hence, we will first analyze the phase stability of the Set-I compounds to see if our DFT calculations are able to  correctly capture the energetics of the known compounds in Set-I and then we will utilize the phase stability data of Set-II compounds to validate the reliability of our approach for the discovery of new stable AMM'Q$_3$ compounds based on prototype decoration through HT-DFT calculations.

\subsubsection{Set-I compounds}
As detailed in the Methods section and as well as in other references\cite{curtarolo2013high,saal2013materials,kirklin2015open,jain2013commentary,curtarolo2012aflowlib, jain2016computational, jain2011high, emery2016high} the hull distance (hd) is a metric of the thermodynamic stability of a compound. If the formation energy of the compound breaks the convex hull, then it is considered to be  thermodynamically stable with hd = 0, indicating the likelihood of its synthesizability. On the other hand, compounds with a small positive hull distance (typically within a few tens of meV/atom) are called metastable and  may also be in some cases experimentally synthesized\cite{sun2016thermodynamic,aykol2018thermodynamic}. According to this criterion, all experimentally known AMM'Q$_3$ compounds should possess zero or small positive hull distances. Our analysis reveals that in Set-I, all (39) Type-I compounds and all but one (29) Type-II compounds have hd = 0, which is in line with our expectations. The one Type-II compound that has a small positive hull distance is Eu$^{2+}$CuEu$^{3}$+S$_3$ (hd = 37 meV/atom). Also, all but three (47) Type-III compounds of Set-I have hd = 0. These three Type-III compounds are CsCoYbS$_3$ (hd =192 meV/atom), CsCoYbSe3 (hd =151 meV/atom), CsZnYbSe$_3$ (hd=76 meV/atom). So, 115 of 119 experimentally synthesized compounds in Set-I are thermodynamically stable in the OQMD. Hence, stability is an excellent metric for the synthesizability of the predicted compounds. From this analysis, we also find that one of the Type-II compounds in Set-I, BaAgErS$_3$, possesses a small positive hd of 12 meV/atom when its calculation is performed on the experimental crystal structure having C2/m SG (\#12), indicating that it is metastable at 0 K in this structure.   Interestingly, our HT-DFT calculations reveal that BaAgErS$_3$ is stable T = 0 K in the KCuZrSe$_3$ structure type (see SI Figure S2). 

\subsubsection{Set-II compounds}
Since the 73 compounds in Set-II did not exist in OQMD before our HT-DFT calculations, we generate their crystal structures through prototype decorations as mentioned before to perform DFT calculations and T = 0 K phase stability analysis. As these compounds have already been synthesized experimentally, our DFT calculations and phase stability analysis provide a key test of our methodology.  It also gives us an opportunity to examine how reliably our calculations can predict hitherto unknown stable AMM'Q$_3$ compounds.  After performing the thermodynamic stability analysis, we found that among the  73 compounds in Set-II (type-I: 1, type-II: 60, type-III: 12), 64 compounds have hd = 0, and only 4 compounds (SrCuYbS3: hd = 84 meV/atom, BaCuYbTe$_3$: hd=62 meV/atom, EuCuYbS$_3$: hd=72 meV/atom, PbCuYbS$_3$: hd =104 meV/atom) of Type-II and 5 compounds (CsMnYbSe$_3$: hd=8 meV/atom, CsZnYbS$_3$ : hd=163 meV/atom, CsZnYbTe$_3$: hd=71 meV/atom, RbZnYbSe$_3$: hd=102 meV/atom, RbZnYbTe$_3$: hd=82 meV/atom) of Type-III have positive hull distances.


\begin{table*}
\begin{center}
\begin{tabular}{ |p{4cm} | p{3cm}| p{3cm} | p{3cm} |  p{3cm} |} 
\hline
	& \textbf{Type-I} &	\textbf{Type-II} & 	\textbf{Type-III} & 	\textbf{Total} \\ [4ex]
\hline	

\textbf{A} &	Li, Na, K, Rb, Cs, In, Tl (7) &	Mg, Ca, Sr, Ba, Ge, Sn, Pb, Eu
(8) &	Li, Na, K, Rb, Cs, In, Tl (7) &	-  \\ [4ex]

\hline
\textbf{M} &	Cu, Ag, Au (3) &	Cu, Ag, Au (3) &	Mn, Fe, Co, Ni, Zn, Cd, Hg, Eu
(8)	& - \\ [4ex]
\hline
\textbf{M'} &	Ti, Mn, Zr, Mo, Ru, Pd, Hf, W, Re, Os, Ir, Pt, Ce
(13) &	Sc, Y, La, Ce, Pr, Nd, Sm, Eu Gd, Tb, Dy, Ho, Er, Tm, Yb, Lu (16) & 	Sc, Y, La, Ce, Pr, Nd, Sm, Eu Gd, Tb, Dy, Ho, Er, Tm, Yb, Lu (16) &	- \\ [4ex]

\hline

\textbf{Q} &	S, Se, Te (3) &	S, Se, Te (3) &	S, Se, Te (3) &	- \\ [4ex]

\hline 
\textbf{Total number of initial compositions} &
	7 $\times$ 3 $\times$ 13 $\times$ 3 = 819 &	8 $\times$ 3 $\times$ 16 $\times$ 3 = 1152 &	7 $\times$ 8 $\times$ 16 $\times$ 3 = 2688 &	4659 \\ [4ex]

\hline
	
\textbf{Thermodynamically stable compounds} &	69  &	231	 & 328 &	628 \\ [4ex]

\hline

\textbf{Thermodynamically metastable compounds}  &	59  &	282	 & 511 & 	852 \\ [4ex]

\hline
\end{tabular}
\caption{\textbf{A table summarizing the design and discovery of new materials through prototype decoration}. Elements chosen for decorating the prototype AMM'Q$_3$ structures and  results of HT-DFT calculations are shown here. The values within the parentheses indicate the total number of elements used in each of three categories of compounds for substitutions at the A, M, M' and Q sites. All generated initial compositions satisfy the charge-neutrality criteria based on their nominal oxidation states. The numbers of predicted stable (hd = 0) and low-energy metastable (0 $<$ hd $\leq$ 50 meV/atom) compounds in the table exclude the experimentally known 192 AMM'Q$_3$ compounds.}
\end{center}
\end{table*}

Thus, our HT-DFT calculations based on the decorated structures successfully capture the stability of all but 13 (out of 192) of the experimentally known AMM'Q$_3$ compounds. The 13 compounds which are experimentally observed but with hd $>$ 0 all contain Eu and Yb, and the calculated energetics of these compounds may originate from the choice of incorrect pseudopotentials as discussed later. These results and analysis give us strong confidence in designing and discovering new AMM'Q$_3$ compounds using the HT-DFT calculations and thermodynamic T=0 K stability as a metric for synthesizable compounds. We have provided the phase stability data of all 192 experimentally known AMM'Q$_3$ compounds in the SI.

\subsection{Novel AMM'Q$_3$ compounds}
After performing the T= 0 K stability analysis of all newly decorated AMM'Q$_3$ compounds for which HT-DFT calculations are performed in all 7 structure types, we discover a large number of 628 (type-I: 69, type-II: 231, type-III:328) thermodynamically stable compounds that exclude the experimentally known 192 compounds. To put this number into perspective,  the OQMD (containing more than 900,000 entries as of September 2020)  has stable (hd = 0) 1161 full-Heuslers (SG: Fm$\bar{3}$m, \#225), 618 half-Heuslers (SG: F$\bar{4}$3m, \#216), 353 cubic (SG: Pm$\bar{3}$m, \#221) perovskites, 242 orthorhombic (SG: Pnma, \#62) perovskites.   Similar to the experimentally known AMM'Q$_3$ compounds, the KCuZrSe$_3$ and Eu$_2$CuS$_3$ structure types are the most common, which constitute 41\% and 25\% of the predicted stable compounds, respectively (Figure 3b). Among the other structure types, NaCuTiSe$_3$ and TlCuTiTe$_3$ are quite common, which constitute 19\% and 9\% of the predicted stable compounds. The rest 6\% compounds have the BaCuLaS$_3$ and BaAgErS$_3$ structure types. However, we found no new stable compound in the Ba$_2$MnS$_3$ structure type.  Our analysis shows that 570 out of 628 compounds possess finite bands that range from 0.2 eV to 5.34 eV, among which most of the compounds have band gaps within 1.5 eV (Figure 3c).  This is not surprising since all the decorated compositions are charge-balanced.  In addition, we found 852 potentially synthesizable  metastable compounds (Type-I: 59, Type-II: 282, Type-III: 511) with small positive hull distances (i.e., 0 $<$  hd  $\leq$ 50 meV/atom).  A summary of the HT-DFT calculations is shown in Table-I, and the lists of all predicted stable and metastable compounds are given in the SI.

We further examine the predicted stable compounds within each type in terms of their chemistries (sulfides, selenides, and tellurides) and the cations (A, M, M').  The results are displayed as bar charts in Figures 4, where the bar corresponding to an element represents the number of stable compounds that contain it. We see that:  (a) the elements that form the stable compounds constitute almost the entire periodic table. (b) There are more Cu-containing compounds in Type-I and Type-II categories compared to Ag or Au. In Type-III, there are more Mn compounds than other M elements with a 2+ oxidation state. (c) The number of stable compounds increases from Type-I (69) to Type-II (231) to Type-III (328). This is not surprising given that the number of elements that can occupy the M and M' sites (satisfying the charge-neutrality criteria) increase from Type-I to Type-III compounds (Table-I). A general trend that is noticeable across three types is that as we go from sulfides to selenides to tellurides, the number of compounds with smaller A cations decrease whereas the number increases with larger A cations. For example, there are two, two, and no Li-containing compounds in sulfides, selenides, and tellurides of Type-I compounds, respectively. Similarly, the number of compounds that have Sr increase from sulfides (7) to selenides (20) to tellurides (34) in Type-II.

\subsection{Lattice thermal transport properties}
We now focus on exploring the lattice thermal transport properties of the predicted stable AMM'Q$_3$ compounds. An accurate estimation of  $\kappa_l$ of a compound within a first-principles DFT framework is computationally very expensive\cite{li2014shengbte}.  Hence, the calculations of $\kappa_l$ for all the predicted stable compounds would require a massive amount of computational resources. However, to demonstrate the thermal transport properties of our newly predicted stable compounds, we randomly select a handful of compounds with some criteria.  We first screen for non-magnetic and semiconducting compounds, where the lattice contribution dominates the total thermal conductivity. Next, we search for those compounds which have the KCuZrSe$_3$ structure type as it possesses the highest crystal symmetry (SG: Cmcm, \#63) and the smallest unit cell (12 atoms).  Finally, we randomly select 10 compounds for $\kappa_l$ calculations. The selected compounds, which include sulfides, selenides and tellurides, are: CsCuZrS$_3$, BaCuScSe$_3$, BaCuScTe$_3$, BaCuTbSe$_3$, BaAgGdSe$_3$, CsZnYS$_3$, CsZnGdS$_3$, CsZnScSe$_3$, CsZnScTe$_3$, CsCdYTe$_3$. We note that this list also includes Type-I (the first in the list), Type-II (the next 4) as well as Type-III (the last 5) compounds. The electronic structures and phonon dispersions of these compounds are given in the SI.

\begin{figure*}
\centering
\includegraphics[scale=0.7, trim={0cm  1.5cm 0cm 0cm}, clip]{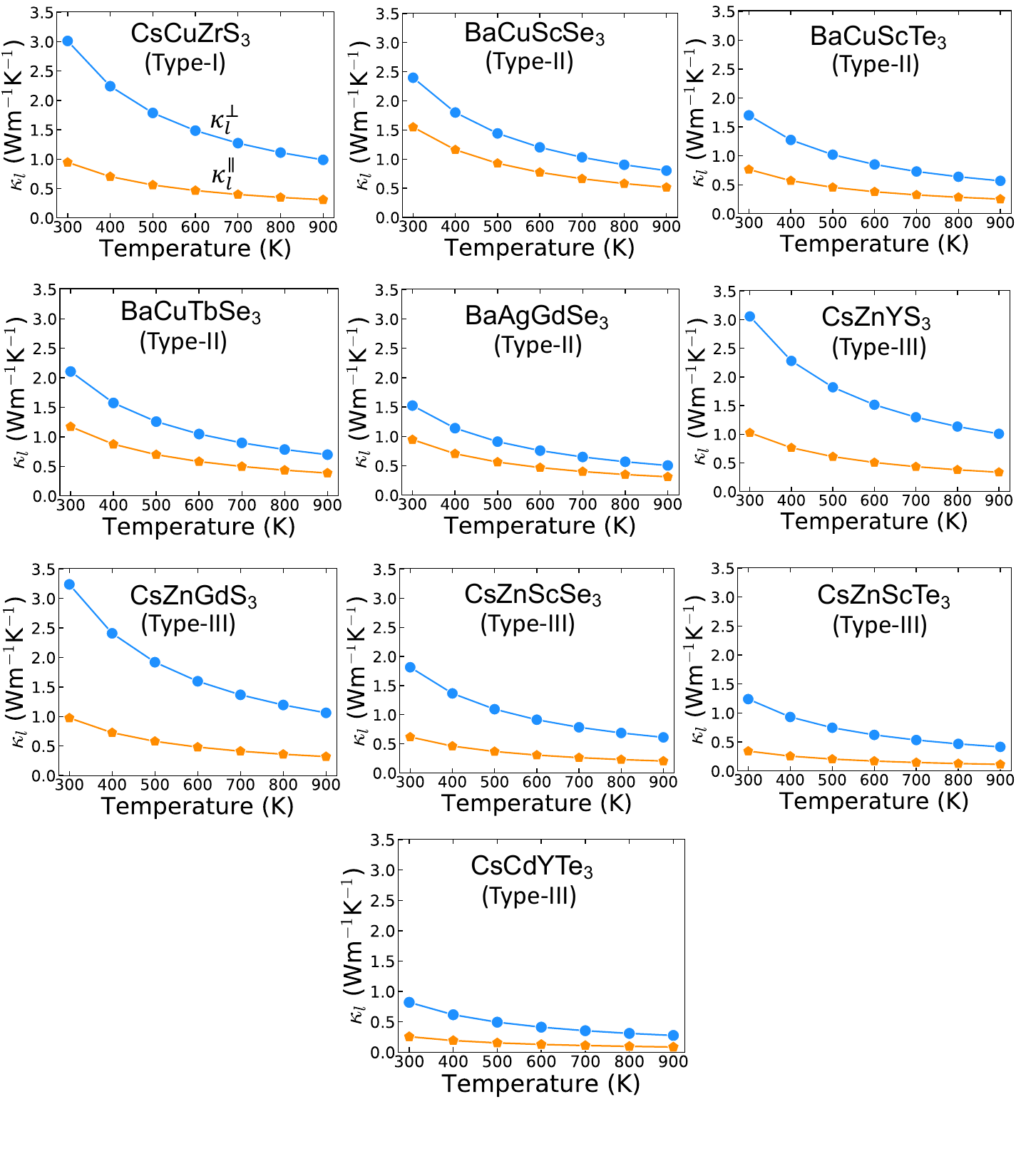}
\caption{\textbf{Lattice thermal transport properties of the AMM'Q$_3$ compounds}. The calculated temperature-dependent lattice thermal conductivity ($\kappa_l$) of 10 predicted stable AMM'Q$_3$ compound which are non-magnetic and semiconducting. $\kappa_l^{\bot}$ (blue disk) and  $\kappa_l^{\Vert}$ (orange pentagon) are the two components of $\kappa_l$ that are perpendicular and parallel to the stacking directions in the crystal structure, respectively.}
\end{figure*}

We calculate the $\kappa_l$ of these 10 compounds using the PBTE (see Methods section) and present the results in Figure 5.  We see that all these compounds exhibit very low $\kappa_l$ where the  in-plane $(\kappa_l^{\bot})$ and the cross-plane $(\kappa_l^{\Vert})$ components are lower than 3 Wm$^{-1}$K$^{-1}$ and 1.2 Wm$^{-1}$K$^{-1}$, respectively,  for T $\geq$ 300 K. Here, $\kappa_l^{\bot}$ is perpendicular to the stacking direction of the layers in the crystal structure of the AMM'Q$_3$ compounds and $\kappa_l^{\Vert}$ is parallel to it. As a reference, we compare our calculated $\kappa_l$ with that of a prototypical thermoelectric material SnSe, which was experimentally shown to possesses  low-$\kappa_l$ that leads to  a very high thermoelectric figure-of-merit\cite{zhao2014ultralow}. The  measured\cite{wu2017direct}  $\kappa_l^{\bot}$ and $\kappa_l^{\Vert}$  for single-crystalline stoichiometric samples of  SnSe  are 1.9 Wm$^{-1}$K$^{-1}$ and 0.9 Wm$^{-1}$K$^{-1}$, respectively, at 300 K, which become much lower in the off-stoichiometric polycrystalline samples\cite{zhao2014ultralow}.  Further examination of the results in Figure 5 reveals that in terms of the anisotropy of the $\kappa_l^{\bot}$ and $\kappa_l^{\Vert}$ components, Type-I and Type-III compounds are quite similar, but Type-II compounds are different from the rest i.e.,  $\kappa_l^{\bot}/\kappa_l^{\Vert}$ (Type-I/III) $>$ $\kappa_l^{\bot}/\kappa_l^{\Vert}$ (Type-II). The difference in the anisotropy of the properties arises from the fact that in Type-II compound the electrostatic attractions between A$^{2+}$ and [MM'Q$_3$]$^{2-}$ layers are stronger than that of between A$^{1+}$ and [MM'Q$_3$]$^{1-}$   layers in Type-I/III compounds. The stronger interlayer interactions give rise to a shorter interlayer distance in Type-II, which make $\kappa_l^{\bot}$ and $\kappa_l^{\Vert}$ less anisotropic.

To gain a deeper understanding, we examine the lattice dynamics and thermal transport properties of BaCuScTe$_3$ (Type-II) and CsCdYTe$_3$ (Type-III) in detail. Our analysis reveals that the underlying physical principles governing the low-$\kappa_l$ in Type-II compounds are different from the Type-I/III compounds. We start our analysis with the phonon dispersion of BaCuScTe$_3$ (Figure 6a) which shows the presence of very low-frequency acoustic ($<$ 45 cm$^{-1}$), and optical ($\sim$ 15 cm$^{-1}$ along X-S-R directions) phonon modes, which give rise to low phonon group velocities. The phonon dispersion and the phonon density of states (Figure 6b) of BaScCuTe$_3$ show that a strong hybridization exists between the phonon branches up to 100 cm$^{-1}$, where Ba and Te atoms have large contributions. Soft acoustic phonon branches give rise to very low sound velocities and a strong hybridization between the phonons at low energies enhances the phonon scattering phase space. Both these factors help suitably to give rise to a very low-$\kappa_l$ ($\kappa_l^{\bot}$ = 1.7 Wm$^{-1}$K$^{-1}$ and $\kappa_l^{\Vert}$ = 0.76 Wm$^{-1}$K$^{-1}$ at 300 K). On the other hand, the phonon dispersion of CsCdYTe$_3$ (Figure 6d) features nearly dispersion-less optical phonon branches along the X-S-R-Z directions in the Brillouin zone at low energies, which are the characteristics of rattler vibrations in the crystal structure. In addition, it also has soft acoustic phonon branches ($<$ 35 cm$^{-1}$). The calculated $\kappa_l$ of CsCdYTe$_3$ becomes ultralow with $\kappa_l^{\bot}$  and $\kappa_l^{\Vert}$ being 0.82 Wm$^{-1}$K$^{-1}$ 0.25 Wm$^{-1}$K$^{-1}$, respectively at 300 K.

Rattler phonon modes are highly localized, which strongly inhibit the transport of phonons, giving rise to ultralow-$\kappa_l$ in many crystalline solids such as TlInTe$_2$\cite{jana2017intrinsic}, CsAg$_5$Te$_3$\cite{lin2016concerted}, etc. It was shown that the filler atoms in clathrates\cite{tadano2015impact} and  skutterudites\cite{li2015ultralow}  act as ideal rattlers, which  give rise to  dispersion-less phonon branches where the phonon frequencies remain highly localized having very small participation ratio (PR) values $\sim$ 0.2 (see Methods section). The PRs of the phonon modes of BaCuScTe$_3$ and CsCdYTe$_3$ are color-coded in Figures 6a and 6d, respectively. We see that while  most of the low-energy phonon modes ($<$ 100 cm$^{-1}$) of BaCuScTe$_3$ have PR values close to 1, signifying the absence of phonon localization, the low-energy dispersion-less phonon branches of CsCdYTe$_3$ have small PRs ( $<$ 0.2), indicating their highly localized nature of the phonon modes. The phonon density of states (Figure 6e) also reveals that  these localized phonons primarily arise from the Cs atoms (confined in 25-55 cm$^{-1}$) that act as rattlers. Thus, our analysis shows that the ultralow-$\kappa_l$ in CsCdYTe$_3$ is primarily caused by the localized vibrations of the rattling phonons.

\begin{figure*}
\centering
\includegraphics[scale=0.50, trim={0cm  0cm 0cm 0cm}, clip]{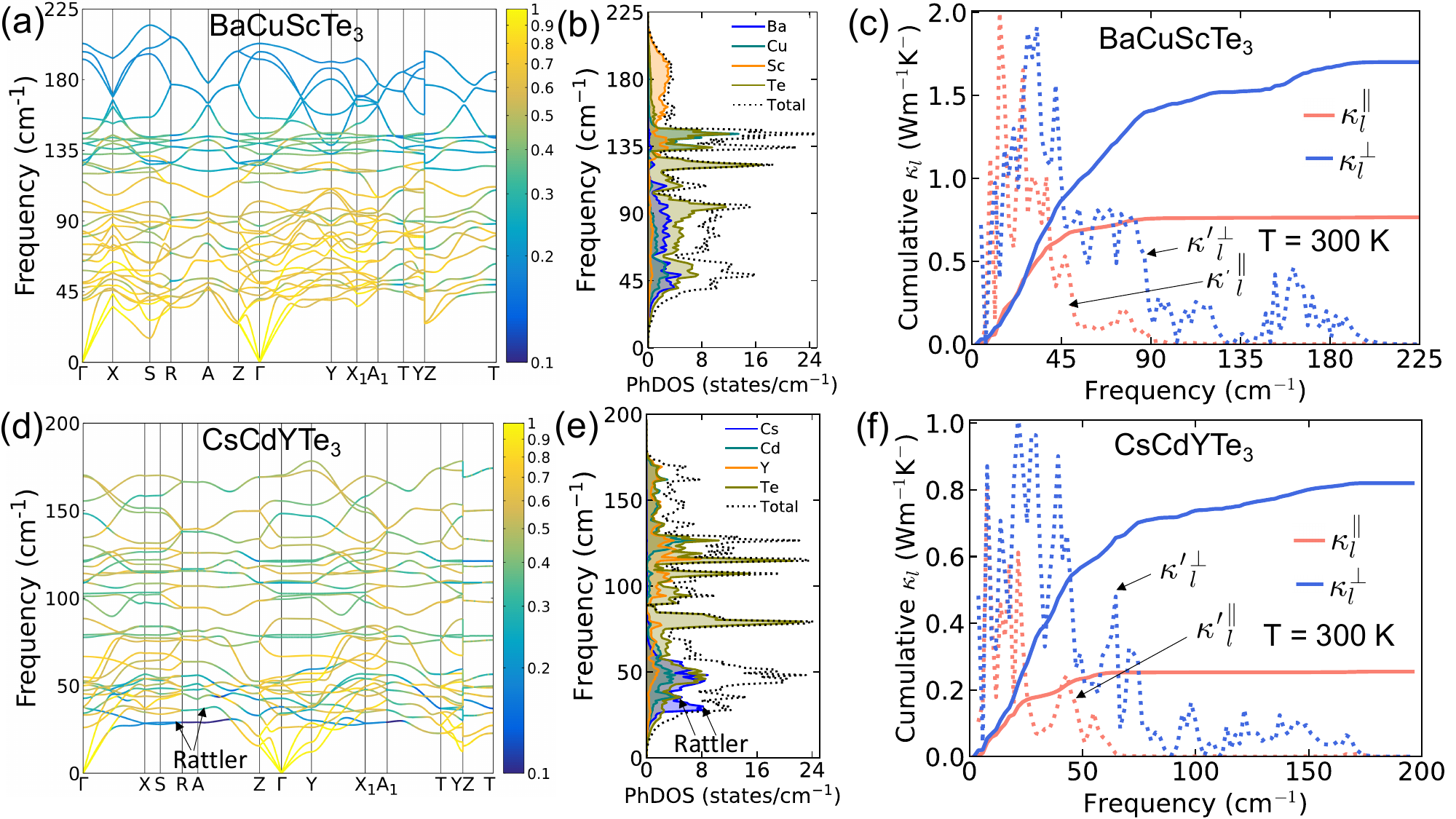}
\caption{\textbf{Harmonic and anharmonic lattice dynamical properties of the AMM'Q$_3$ compounds.} Harmonic phonon dispersions, atom-resolved phonon density of states, anharmaonic frequency-cumulative  $\kappa_l$, and their derivatives with respect to the frequency for BaCuScTe$_3$ (a-c) and CsCdYTe$_3$ (d-f). The phonon dispersions are color coded with the participation ratios of the phonon modes (indicated in the color bar).  $\kappa_l^{\bot}$ (solid blue line) and $\kappa_l^{\Vert}$ (solid red line) are perpendicular and parallel to the stacking directions in the crystal structure of the AMM'Q$_3$ compound, respectively. ${\kappa'_l}^{\bot}$ (dotted blue line) and ${\kappa'_l}^{\Vert}$ (dotted red line) are the first-order derivatives of $\kappa_l^{\bot}$ and  $\kappa_l^{\Vert}$, respectively, with respect to the frequency, which are in arbitrary units in the figures.}
\end{figure*}

We further investigate the origin of such poor thermal transport properties in BaCuScTe$_3$ in terms of a more fundamental material quantity, the lattice anharmonicity.  Strong lattice anharmonicity is one of the important factors that induces very low-$\kappa_l$   in compounds like SnSe\cite{li2015orbitally}, NaSbTe$_2$\cite{nielsen2013lone}, etc. To estimate the strength of the intrinsic anharmonicity,  we calculate the   macroscopic average Gruneisen parameter, $\gamma = \sum_i{\gamma_i}C_{v,i}/\sum_iC_{v,i}$, where $\gamma_i$ and $C_{v,i}$ are the Gruneisen parameter and specific heat capacity at constant volume  for the $i$-th phonon mode. The calculated $\gamma$ of BaCuScTe$_3$ (1.5) is larger than that of CsCdYTe$_3$ (1.2), signifying the presence of stronger anharmonicity in the former. These $\gamma$ values are comparable to that of NaSbSe$_2$ (1.7), NaSbTe$_2$ (1.6), NaBiTe$_2$ (1.5) etc. compounds which are experimentally\cite{nielsen2013lone} shown to  possess ultralow-$\kappa_l$. Thus, we see that while the low-$\kappa_l$   in Type-II compounds is caused by the low-sound velocities and a stronger lattice anharmonicity, the presence of rattler cations in Type-III as well Type-I compounds (see SI Figure S4) are primarily responsible for inducing low-$\kappa_l$  in them.  

To examine which phonons are primarily responsible for the conduction of heat in BaCuScTe$_3$ and CsCdYTe$_3$, we plot the cumulative-$\kappa_l$ and their first-order derivatives ($\kappa'_l$) with respect to the frequency at T = 300 K.   We see from Figure 6c that while the acoustic and low-energy optical phonons up to 90 cm$^{-1}$ mainly contribute to  $\kappa_l^{\bot}$,    $\kappa_l^{\Vert}$  is primarily contributed by the phonons up to  45 cm$^{-1}$ in BaCuScTe$_3$.  On the other hand, Figure 6f shows that while the acoustic as well as the optical phonons up to 100 cm$^{-1}$ have large contributions towards $\kappa_l^{\bot}$, only the acoustic phonons (up to 35 cm$^{-1}$) primarily carry heat for 
$\kappa_l^{\Vert}$ in CsCdYTe$_3$. We also notice that the anisotropy $(\kappa_l^{\bot}/\kappa_l^{\Vert}$ = 3.3 at T = 300 K) in CsCdYTe$_3$ is much larger that of BaCuScTe$_3$ ($\kappa_l^{\bot}/\kappa_l^{\Vert}$ = 2.2 at T = 300 K). The origin of this anisotropy can be attributed to the contrasting interlayer and intralayer interactions in BaCuScTe$_3$ and CsCdYTe$_3$.  For example, the analysis of the interatomic force-constants (IFCs) reveals that interlayer interactions in CsCdYTe$_3$ are much weaker (IFC$_{(Cs-Te)}$ = -0.333 eV/$\AA^2$)  compared to BaCuScTe$_3$ (IFC$_{(Ba-Te)}$ =-1.204 eV/$\AA^2$), which makes the transport of optical phonons (above 25 cm$^{-1}$) along the stacking direction (i.e., $\kappa_l^{\Vert}$) of CsCdYTe$_3$ less effective. On the other hand, the intralayer interactions in CsCdYTe$_3$ are much stronger (IFC$_{(Cd-Te)}$= -3.923 eV/$\AA^2$, IFC$_{(Y-Te)}$: -2.216 eV/$\AA^2$) than those of BaCuScTe$_3$ (IFC$_{(Cu-Te)}$= -2.396 eV/$\AA^2$, IFC$_{(Sc-Te)}$= -1.890 eV/$\AA^2$). As a result, while phonons up to 90 cm$^{-1}$ mainly carry the heat for  $\kappa_l^{\bot}$ in BaCuScTe$_3$, in CsCdYTe$_3$ they are carried by phonons with frequencies up to 100 cm$^{-1}$ very effectively.

\section{Discussion}
 
We notice that in Set-I and Set-II, all the known compounds that have positive hull distances have either Yb$^{3+}$ or Eu$^{3+}$ cations in them. Given the fact OQMD used Yb\_2 and Eu\_2 PPs which are intended for compounds having Yb$^{2+}$ and Eu$^{2+}$ cations, those energetic results are somewhat suspect. However, we note that our HT-DFT calculations predict the stability of other rare-earth elements containing known compounds in this family correctly. Hence, the newly predicted compounds containing Eu$^{3+}$ and Yb$^{3+}$ should be taken with caution. We note that none of our predicted stable compounds contain these cations. However,  11 and 7 of the predicted metastable compounds contain Eu$^{3+}$ and Yb$^{3+}$ cations, respectively.

Concerning the experimental validation of our prediction, we note that the experimental synthesis of a large number of compounds is a daunting task. While automated synthesis of a large number of compounds is  now possible through high-throughput experimental facilities\cite{green2017fulfilling}, here, we suggest only four compounds (Table II) that can be immediately picked up for experimental verification of our prediction. These compounds contain no toxic element and their calculated lattice thermal conductivity is very low. In Table II, we provide their DFT calculated band gaps, $\gamma$, and average  $\kappa_l$ calculated at 300 K. These calculated quantities along with $\kappa_l$ can be compared with the experimentally measured values of these materials. Finally, we note that $\kappa_l$ for each compound has been calculated using only three-phonon scattering processes. The inclusion of additional four-phonon scattering rates\cite{xia2020particlelike,xia2020high} and  grain-boundary\cite{pal2020microscopic} limited phonon scattering mechanisms could further lower the calculated $\kappa_l$ in this family of compounds. Also, as the electronic structure of the compounds features nearly flat bands and multiple peaks near the valence/conduction band extrema (Figure S3, SI), some of these compounds are expected to exhibit good potential for thermoelectric applications as well. 


\begin{table}
\begin{center}
\begin{tabular}{ | m{2cm} | m{2cm}| m{2cm} | m{2cm} |} 
\hline
Compound &	Band gap (eV) & Average	$\gamma$ at 300 K &	Average $\kappa_l$ (Wm$^{-1}$K$^{-1}$) 
at 300 K \\

\hline
BaCuScTe$_3$ &	0.39 &	1.51 & 	1.23 \\
\hline
BaAgGdSe$_3$ &	1.45 &	1.60 &	1.24 \\
\hline
CsZnScTe$_3$ &	1.20 &	1.30 &	0.79 \\
\hline
CsCdYTe$_3$ &	1.74 &	1.30 &	0.54 \\

\hline
\end{tabular}
\caption{\textbf{Suggested compounds for experimental synthesis and characterization}. A list of 4 predicted stable compounds are suggested for the experimental synthesis and measurement of their properties. The DFT-calculated band gap, macroscopic average Gruneisen parameter ($\gamma$), the average $\kappa_l$ are provided to compare against experiments.}
\end{center}
\end{table}

In summary, we use HT-DFT calculations to discover a large number of 628 thermodynamically stable quaternary chalcogenides (AMM'Q$_3$). As all compositions in this family are charge-balanced, our analysis shows that 570 of 628 compounds possess finite band gaps which vary between 0.2 and 5.34 eV.  Our calculations of the thermal transport properties show that AMM'Q$_3$ compounds exhibit intrinsically very low $\kappa_l$, and the anisotropy in $\kappa_l$ is much smaller in Type-II compounds compared to Type-I/III compounds.  Our analysis further reveals that low-$\kappa_l$ in this family originates either due to the presence of rattling cations (in Type-I/III compounds) or stronger lattice anharmonicity (Type-II compound). While the rattler cations give rise to localized phonon modes that inhibit the propagation of phonons, a stronger lattice anharmonicity enhances the phonon scattering phase space, leading to a low-$\kappa_l$. In addition, there exists a strong coupling between the acoustic and low-energy optical phonon modes in Type-II compounds which increases the phonon scattering rates of the heat-carrying phonons. Our work is thus interesting not just from the perspective of new materials discovery but also for finding the presence of low $\kappa_l$ in them, which hold promises for further research and possible applications in energy materials and related devices.

\section{Methods}

\subsection{DFT calculations}
We performed DFT calculations using the Vienna Ab-initio Simulation Package (VASP)\cite{kresse1996efficiency} using the projector-augmented wave (PAW)\cite{kresse1999ultrasoft} potentials with the Perdew-Burke-Ernzerhof (PBE)\cite{perdew1996generalized} generalized gradient approximation (GGA) to the exchange-correlation functional. The atomic positions and other cell degrees of freedom of the compounds were fully relaxed and spin-polarized calculations were performed for compounds that contain partially filled d or f-shells elements with a ferromagnetic arrangement of spins in accordance with the high-throughput framework as laid out in the qmpy suite of tools\cite{saal2013materials,kirklin2015open}. For more details on the calculation parameters, we refer to Ref.\cite{saal2013materials,kirklin2015open}. T=0 K phase stability analysis often serves as an excellent indicator for the possibility of synthesizability of a predicted compound in the laboratory\cite{zakutayev2013theoretical,gautier2015prediction,anand2019double}. To assess the T = 0 K thermodynamic stability of the compounds, we calculate their formation energies ($\Delta$H$_f$) utilizing the DFT total energy (ground state) of each compound using the formula, $\Delta$H$_f$($\sigma$) = E($\sigma$) - $\sum_i$ n$_i\mu_i$, where E is the DFT total energy (at 0 K) of an AMM'Q$_3$ compound in a crystal structure denoted by $\sigma$, $\mu_i$ is the chemical potential of element $i$ with its fraction n$_i$ in that compound. For each composition, we used a number of prototype crystal structures, $\sigma$, based on structural prototypes of known AMM'Q$_3$ compounds. To determine the thermodynamic stability of a compound, we need to compare its formation energy against all its competing phases, not only against other compounds at the same composition. To this end, we generated the quaternary phase diagram (i.e., the T = 0 K convex hull of the A-M-M'-Q phase space) for every AMM'Q$_3$ compound considering all elemental, binary, ternary, and quaternary phases present in the OQMD, which (as of September 2020) corresponded  to nearly 900,000 entries of DFT-calculated energies. The calculated convex hull distance (hd, explained in the next section) then serves as a metric to determine whether a compound is stable (i.e.,  hd = 0), metastable (small positive hd), or unstable (large positive hd).

\begin{figure*}
\centering
\includegraphics[scale=0.50, trim={0cm 2cm 0cm 0cm}, clip]{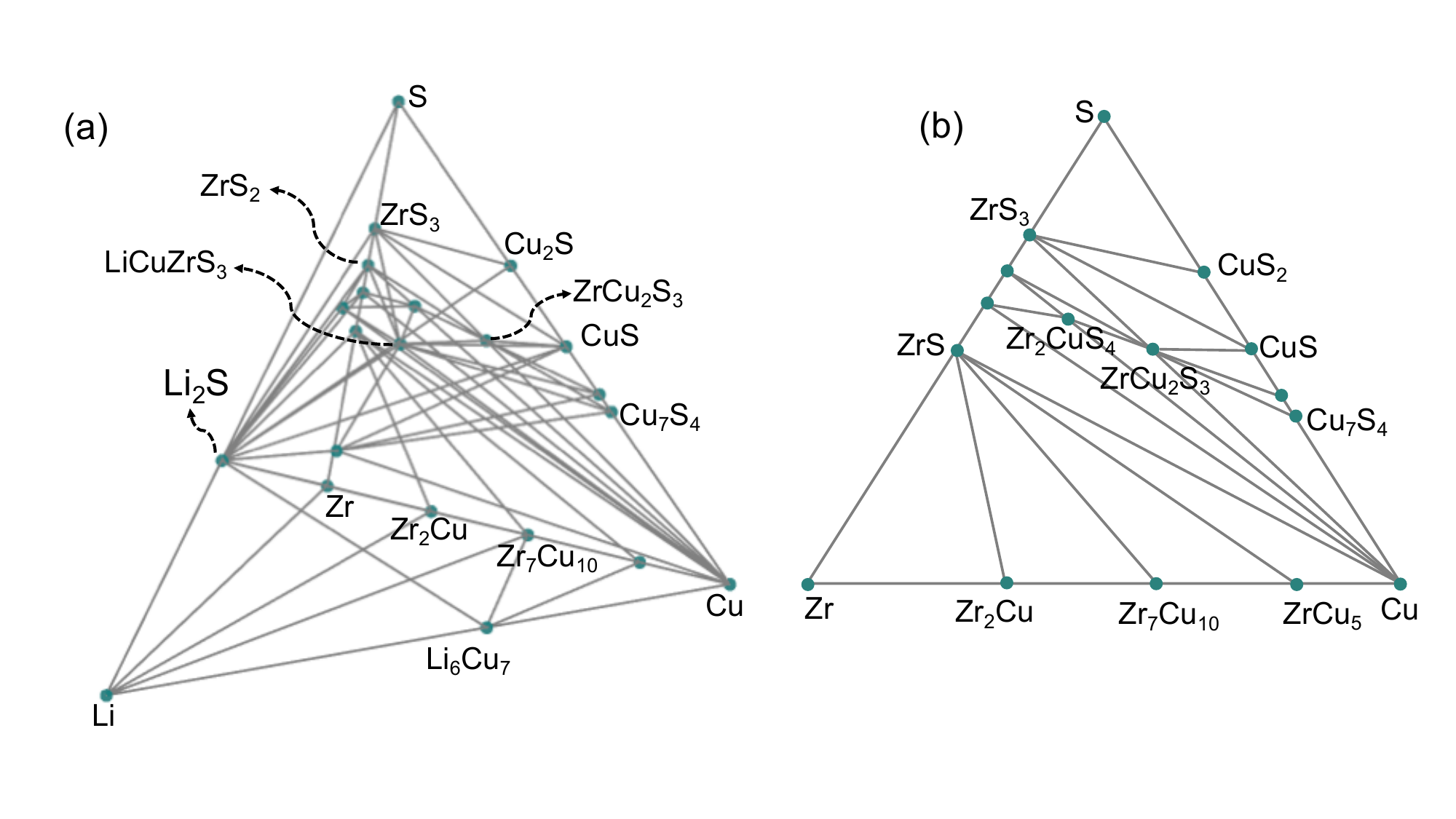}
\caption{\textbf{Visualization of quaternary and ternary phase diagrams}. (a) The four dimensional phase diagram (T= 0 K) of  the Li-Cu-Zr-S quaternary system, which is the isometric shot of a Gibbs' tetrahedron.  (b) One of the faces of the tetrahedron represents the three-dimensional phase diagram of the Cu-Zr-S ternary system which is presented as the Gibbs' triangle.  Each cyan node in (a) and (b) represents a stable compound. For clarity, we do not show any metastable/unstable compounds and mark only few stable compounds in this figure. LiCuZrS$_3$ is one of the predicted stable compounds in this work, which and its competing phases are denoted with dashed arrows in (a).}
\end{figure*}

\subsection{Convex hull construction}
To construct the convex hull of an AMM'Q$_3$ compound, it is necessary to identify the set of phases (elemental, binary, ternary as well as quaternary) in the four-dimensional composition space of A-M-M'-Q that have the lowest formation energies at their compositions. In OQMD, we present the convex hull of a quaternary compound through a four-dimensional phase diagram which is represented as an isometric shot of the Gibbs' tetrahedron (Figure 7a). Each face of the tetrahedron represents a three-dimensional phase diagram of a ternary composition that is represented as the Gibbs' triangle (Figure 7b). The vertices in Figures 7a \& 7b represent the  elements constituting the quaternary and ternary phase space, respectively, and the edges connecting any two vertices represent the binary composition axis between those two elements.   Any node within Figures 7a \& 7b represents a stable compound, which is denoted by a cyan disk. The metastable and unstable compounds are not shown in these figures for clarity which fall off the nodes. The stability of an AMM'Q$_3$ compound is given by the difference (which is defined as the hull distance, hd = $\Delta$H$_f$  - $\Delta$H$_e$) between the calculated formation energy ($\Delta$H$_f$) of an AMM'Q$_3$ compound under consideration and its hull energy ($\Delta$H$_e$) at that composition. The hull energy is defined as the energy at the convex hull  at that AMM'Q$_3$ composition. By definition, the hd of a stable compound is zero, whereas for metastable and unstable compounds they are real positive numbers. In keeping with the heuristic conventions used in literature\cite{zakutayev2013theoretical, cerqueira2015identification, wu2013first} we term those AMM'Q$_3$ compounds to be metastable whose hull distances lie within 50 meV/atom above the convex hull (i.e., 0 $<$ hd $\leq$ 50 meV/atom). The metastable compounds are also potentially synthesizable in the laboratory \cite{sun2016thermodynamic,aykol2018thermodynamic}. 

\subsection{Phonon participation ratio}
Phonon dispersions have been calculated using 2 $\times$ 2 $\times$ 1 supercell of the primitive unit cell using Phonopy\cite{phonopy}. The high symmetry paths in the Brillouin zones were adopted following the conventions used by Setyawan et al.\cite{setyawan2010high}. To examine the extent of localization of the phonon modes, we calculate their phonon participation ratio using the formula\cite{tadano2015impact, pailhes2014localization}: $P(\omega_q) =  (\sum_i^N \frac{{{|e}_i(\omega_q)|}^2}{M_i})^2/\sum_i^N \frac{{{|e}_i(\omega_q)|}^4}{M_i^2}$, where  $e_i(\omega_q)$ is the eigenvector of the phonon mode at q with frequency $\omega$,   $M_i$  is the mass of the $i$-th atom in the unit cell containing a total number of $N$ atoms. The value of $P(\omega_q)$ ranges between 0 and 1. In an ordered crystal, when $P(\omega_q)$ becomes close to 1, it indicates that the phonon mode is propagative where all the atoms in the unit cell participate. On the other hand, very low values of PR  ($\sim$ 0.2)\cite{pailhes2014localization, beltukov2013ioffe} indicate  the strong localization of the  phonon modes (e.g., rattling phonons) where only a few atoms in the unit cell participate in the vibrations. Examples of rattler atoms containing compounds include filled clathrates (Ba$_8$Si$_{46}$ and Ba$_8$Ga$_{16}$Ge$_{30}$)\cite{tadano2015impact,pailhes2014localization}, where the filler atoms act as ideal rattlers that induce ultralow-$\kappa_l$ in them.

\subsection{Thermal conductivity calculations}
We calculate the $\kappa_l$ utilizing the phonon lifetimes obtained from the third-order interatomic force constants (IFCs)\cite{chaput2011phonon, togo2015distributions,zhou2014lattice}, which was shown to reproduce $\kappa_l$  within 5 \% of the experimentally measured $\kappa_l$ in this AMM'Q$_3$ family of compounds\cite{hao2019design, pal2019intrinsically}. We constructed the third-order IFCs of each compound based on DFT calculations of displaced supercell configurations by limiting the cut-off distance (r$_c$) up to the third nearest neighbor. We used 2 $\times$ 2 $\times$ 1 supercell (containing 48 atoms) of the primitive unit cell (with 12 atoms) using thirdorder.py\cite{li2014shengbte} utility for the calculation of IFCs. Using the second and third-order IFCs in the ShengBTE code\cite{li2014shengbte}, we calculate the temperature-dependent phonon scattering rates and $\kappa_l$ utilizing a full iterative solution to the PBTE for phonons using a 12 $\times$ 12 $\times$ 12 q-point mesh. The calculated $\kappa_l$ generally depends on the r$_c$ which accounts for the maximum range of interaction in the third-order IFCs\cite{li2015orbitally}. It was shown that good convergence of $\kappa_l$  was obtained by limiting r$_c$ to the third nearest neighbor within the crystal structure in this family of compounds\cite{pal2019intrinsically}.

\section{Data availability} The data that supports the findings of the work are in the manuscript and Supplementary Information. The structures and energetics of the predicted compounds would be made available through the Open Quantum Materials database (OQMD)  in a future release. Additional data will be available upon reasonable request.

\section{Code availability} Open-source codes are used throughout this work.

\section{Materials \& Correspondence}
Request for materials and correspondence should be addressed to K.P. or C.W.

\section{Acknowledgements} 
\begin{acknowledgements}
K.P. and C.W. acknowledge support from the U.S. Department of Energy under Contract No. DE-SC0014520 (thermal conductivity calculations) and the Center for Hierarchical Materials Design (CHiMaD) and from the U.S. Department of Commerce, National Institute of Standards and Technology under Award No. 70NANB14H012 (HT-DFT calculations).  J.S. and J.H. acknowledges support from the National Science Foundation through the MRSEC program (NSF-DMR 1720139) at the Materials Research Center (phase stability).  Y.X. acknowledges support from Toyota Research Institute (TRI) through the Accelerated Materials Design and Discovery program (lattice dynamics). Y.L. and M.G.K. were supported in part by the National Science Foundation Grant DMR-2003476. K.P. sincerely thanks Sean Griesemer for useful discussion on the abundance of various crystallographic prototypes in the OQMD. We acknowledge the computing resources provided by (a) the National Energy Research Scientific Computing Center (NERSC), a U.S. Department of Energy Office of Science User Facility operated under Contract No. DE-AC02-05CH11231, (b)  Quest high-performance computing facility at Northwestern University which is jointly supported by the Office of the Provost, the Office for Research, and Northwestern University Information Technology, and (c)  the Extreme Science and Engineering Discovery Environment (National Science Foundation Contract ACI-1548562).
\end{acknowledgements}

\section{Author contributions}
K.P. conceived and designed the project. K.P. performed calculations and analysis with help and suggestions from Y.X., J.S., J.H., Y.L, M.G.K, and C.W. C.W. supervised the whole project. All authors discussed the results, provided comments, and contributed to writing the manuscript.

\section{Competing interests} The authors declare no competing financial or non-financial interests.



%

\end{document}